# Accessible Astronomy: Policies, Practices, and Strategies to Increase Participation of Astronomers with Disabilities

**Authors:** Alicia Aarnio[a], Nicholas Murphy[b], Karen Knierman[c], Wanda Diaz Merced, Alan Strauss[d], Sarah Tuttle[e], Jacqueline Monkiewicz[c], Adam Burgasser[f], Lia Corrales[g], Mia Sauda Bovill[h], Jason Nordhaus[i], Allyson Bieryla[b], Patrick Young[c], Jacob Noel-Storr[j], Jennifer Cash[k], Nicole Cabrera Salazar[l], Hyunseop Choi[m]

**Affiliations:** a) University of North Carolina Greensboro, b) Center for Astrophysics | Harvard & Smithsonian, c) Arizona State University, d) University of Arizona, e) University of Washington, f) UC San Diego, g) University of Michigan, h) Texas Christian University, i) Rochester Institute of Technology, j) InsightSTEM, k) South Carolina State University, l) Movement Consulting, m) University of Oklahoma

## 0. Abstract

One outcome of the 2015 Inclusive Astronomy conference was the establishment of an accessibility/disability advocacy group within professional, US-based astronomy, organized by a coalition of disabled astronomers and allies and is supported by the American Astronomical Society (AAS). While the Working Group on Accessibility and Disability (WGAD[1]) has focused on AAS-led initiatives to increase the accessibility of publications, databases, and professional meetings, there is an urgent need to expand these accessibility efforts beyond the professional society and into the wider astronomical community. Our long-term goals include proactively designing learning and working environments to be as accessible as possible, the removal of existing physical, technological, and pedagogical barriers to access, and provision of greater support for the career progress, promotion, and retention of disabled astronomers and educators. Progress toward these goals can be made by establishing and then sustaining a culture of inclusion in which all identities and intersections of identity are equally represented, while recognizing that progress which liberates one identity group may not liberate another in the same way. In the decades since the passage of the Americans with Disabilities Act (ADA), it has become clear that academic departments and research institutions will only undertake the necessary cultural and infrastructure changes if motivated by clear guidelines from funding organizations or ADA non-compliance lawsuits[2,3,4].

In this white paper, we outline the major barriers to access within the educational and professional practice of astronomy. We present current best practices for inclusivity and accessibility, including classroom practices, institutional culture, support for infrastructure creation, hiring processes, and outreach initiatives. We present specific ways—beyond simple compliance with the ADA—that funding agencies, astronomers, and institutions can work together to make astronomy as a field more accessible, inclusive, and equitable. In particular, funding agencies should include the accessibility of institutions during proposal evaluation, hold institutions accountable for inaccessibility, and support efforts to gather data on the status and progress of astronomers and astronomy students with disabilities.

## I. Introduction

Federal law varies in its definition of disability depending on context, but broadly defines disability as "a physical or mental impairment that substantially limits one or more major life activities" and someone who is disabled is "regarded as having such an impairment" and "has a record of the impairment.[5]" Extensive lists are maintained detailing specific conditions, whether they are considered disabling, and what accommodations are available[6]. In this document, we refer to any impairment that, due to the nature of the impairment and/or barriers encountered, disables an individual from participating in astronomy as fully as someone who is able-bodied, neurotypical, and in good mental health. While this document places these groups under the umbrella of disability, individuals within these groups may or may not identify themselves as disabled. Which individuals identify as disabled and which do not is highly nuanced and extremely personal. Deference should always be made to the terms an individual uses. Fundamentally, accessibility is a human right: lack of accessibility constitutes discrimination in and of itself.

Due to a dearth of data specific to astronomy, we refer to general labor force and educational statistics and later discuss what specifically we know about astronomers with disabilities. Data from the United States Census Bureau indicate that in 2014, when surveying for a broad range of disabilities, 27.2% of people living in the US had a disability[7]. Within employment generally, it is well established that people with disabilities are underrepresented (the US Bureau of Labor Statistics reports that in 2018, 19.1% of Americans with disabilities were employed, versus 65% of Americans without disabilities[8]). The lack of representation is in even sharper contrast in academic paths. Studies of Science, Technology, Engineering, and Mathematics (STEM) students' paths found a dramatic decline in representation of people with disabilities: 9-10% of students at the undergraduate level, 5% at the graduate level, and 1% at the doctorate level (Fig. 1)[9]. Multiple recent workplace studies have found that up to 34% of employees surveyed would fit current federal definitions of being disabled[10,11]. Of those, only one third would disclose this to their employer and even fewer would disclose to their colleagues[10]. A majority of these workers report they are not advancing in their careers, and almost half feel their ideas and contributions are not valued[11]. In academia, given that research indicates disabled doctoral students are less likely to be awarded research assistantships (16.4% vs. 24.4%) than their non-disabled counterparts[12] and that STEM employment rates show that people with disabilities are far less likely to be employed than their non-disabled counterparts (21% vs 70%[13]), unwillingness to disclose and a sense of feeling unvalued are perhaps unsurprising. Non-disclosure removes the possibility of requesting needed accommodations, which means barriers to job performance remain in place.

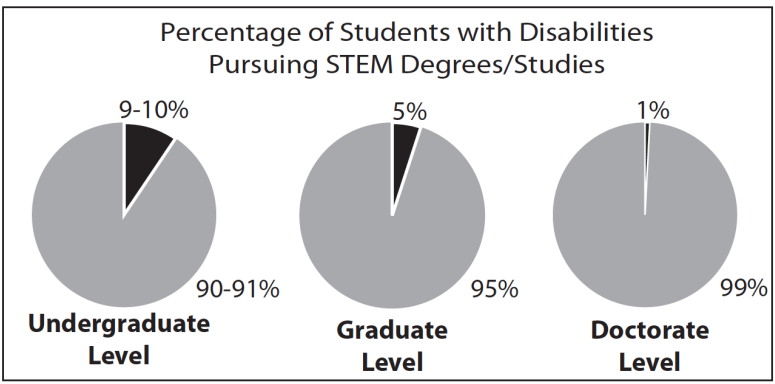

Figure 1. An illustration of loss of students with disabilities over sequential career stages. *Figure reproduced from Moon et al. (2012).*

Specific to astronomy, in the latest AAS membership demographic survey, 3.2% of respondents identified as being deaf or having serious difficulty hearing, being blind or having serious difficulty seeing, or having serious difficulty walking or climbing stairs[14]. While some chose not to answer, overall, 94% of respondents selected "none of the above." This result was interpreted as meaning the majority of the membership is not disabled, but the question is extremely limited in which disabilities it asks respondents about: we are still lacking much information. Furthermore, the career experiences of people with congenital disabilities versus those who became disabled later in life can be very different. Further, data surveying the AAS membership is distinct and discrepant with NSF surveys[14,15], so there is not a clear picture of the reality of astronomers with disabilities' participation in the field; an independent and inclusive assessment is needed. The lack of data is a global phenomenon: the Zero project, conducting global studies regarding social indicators of accessibility, recognized that accessibility and disability demographic data often simply do not exist, so instead they have surveyed whether data on compliance with United Nations accessibility standards even exist. In many cases for the US and other countries, there are no data at all[16].

All of the statistics summarized above serve to shed some light onto the underrepresentation of astronomers with disabilities, even as we lack a complete data set. Within astronomy, we have a census of many aspects of astronomers' identities, but there has not yet been any assessment of the presence, retention, quality of experiences, and progress of people with disabilities. Further, we have no information on aspects of disability beyond senses of sight, hearing, and mobility, as gathered from AAS membership demographic surveys. *Fundamentally, this lack of data means we cannot even begin to grasp of the full extent of the loss to our field due to barriers excluding people with disabilities from participating fully in astronomy.* At the same time, the focus on categorizing individuals into groups defined by a medical impairment does not fully illuminate the accessibility barriers that exist in our profession, and it risks leaving out invisible disabilities. The current approach comes with the risk of diverting attention from needed systemic improvements to maintaining disability as an individual "problem" rather than a collective issue of access and underrepresentation. What follows focuses on the academic and scientific "Astronomical Environment" rather than accommodating specific individuals/conditions.

## II. Barriers to access

*"Barriers to access are conditions or obstacles that prevent individuals with disabilities from using or accessing knowledge and resources as effectively as individuals without disabilities."*[17] Despite passage of the ADA in 1990, many barriers to access still currently exist and prevent the participation of astronomers with disabilities in research and activities necessary for advancement in the field. Each of these barriers, many experienced simultaneously and some sequentially, compound to the point of insurmountability.

Reports indicate less than ~40% of students with disabilities obtain college degrees; the data indicate outcomes tend to be slightly better at two-year institutions than four-year[18]. An estimated 86% of students with psychiatric disabilities withdraw from college before completing their degrees[19]. For graduate students in particular, a significant contributing factor is accessibility offices on campuses which are equipped for helping undergraduates, but largely unprepared to provide support for graduate students[20]. Graduate students often seek informal accommodations, which require disclosure to a professor or advisor[21]. At all educational levels, accommodation provision may not be sufficient due to underfunding and lack of access to resources; supports available vary widely across institutions, and entirely depend on the motivation of the school (which may well be to avoid ADA noncompliance lawsuits). Guides for students with disabilities include advising students to seek a university that provides accommodations that fit the student's needs[22]. Beyond ADA compliance and availability of accommodations, academic culture itself can present barriers: the current metrics of academic competition and evaluation equate humans with test scores, effectively dehumanizing students in the process. Equating value with performance leads to underperformance due to impairment being interpreted as lack of talent or skill, when it should be attributed to accommodation failure. Even when accommodations are available and requested, due to the competitive culture of academia and general stigma surrounding disability, the requests may be scoffed at as invalid, unnecessary, or as attempts by some students to "cheat the system," leading to a reduction in academic standards[23].

Once a disabled student manages to succeed and graduate, pursuing employment presents additional barriers. The academic hiring process is presently subject to much bias and nonuniformity. Within society broadly, intense stigma and negative stereotypes about people with disabilities touch all aspects of our lives. When applying for jobs, people with disabilities experience unfair evaluation that then serves as a barrier to progress. Current metrics and models of productivity feed into expectations that are unrealistic for all, but even more so for astronomers who may need more time to complete work, require rest, or experience periodic hospitalization or flare-ups related to their conditions. "Failing" to live up to high expectations of productivity feeds into many negative stereotypes; this is exacerbated for people with disabilities. The inflexibility and accepted (often celebrated) imbalance of academic work and life is exclusionary in and of itself; this culture is profoundly toxic for people with disabilities.

At the most basic level, astronomers with disabilities experience a general lack of access to information. Publications are often inaccessible to assistive technologies, both in terms of reading papers and submitting manuscripts. In a field as competitive as astronomy, there is no support for astronomers with disabilities to be able to recuperate from the lack of access to information caused by the inability of astronomy-serving institutions to provide it. For one specific example, the International Astronomical Union (IAU) has been in existence for over a century and only just recently recognized that other sensory modes (e.g., sonification) are as equally valid for data analysis as historically employed methods[24,25].

Astronomers with disabilities are often unable to participate in critical social aspects of professional activities. Inaccessible spaces and activities exclude participation, leading to isolation and further hindering progress. Some examples of inaccessible places or events include: restaurants with stairs or only high tables, strobe lighting at a club, alcohol-centric outings, venues filled with smoke or other strong fragrances, spread-out conference venus that necessitate traversing ¼ mile in a short interval between talks, high impact sports for academic discussion groups or conference outings, and loud/overcrowded events.

In the sociopolitical model of disability[26,27], it is not an individual's impairment alone that leads to lack of access and opportunity, but rather the design of social and systemic barriers that exclude people with disabilities. While the social/sociopolitical models of disability have led to much progress in improving the accessibility of public spaces, this model fails to fully capture the experiences of people with disabilities[28]. The Nordic model of disability strikes a balance between barriers that are experienced socially and impairments that are experienced individually[29,30,31]. Even while we dismantle environmental barriers to access, we must remember that an astronomer with a disability will still face work and life impacts due to their impairment, and support can still be provided both for performance at the job of astronomy and for the humans doing the research.

# III. Inequity of experience

Barriers to access, employment structures, and exclusionary practices that can be directly identified and addressed are just part of the ensemble of contributions to an overall much poorer quality of experience for disabled astronomers and students. These structural aspects are potentially more straightforward to identify and address than the critical interpersonal engagement that attends training and professional development. Due to science being a human endeavor, we rely on relationships with advisors, mentors, and colleagues; inability to form these relationships are social barriers that can be detrimental to the point of driving disabled astronomers out of the field at high rates. Some examples are poor mentoring, harassment, bullying, exclusion, negative workplace climate, ableism, and disparaging remarks/attitudes from peers.

Disability is highly stigmatized in our society[32,33]. This stigma is present in astronomy[34,35], where professors often view disability accommodations as an attempt to gain an unfair advantage.[36] Astronomers with invisible disabilities may hide this aspect of their identities for fear that it may prevent them from being accepted to graduate school, getting hired as a postdoc, or being awarded tenure. The stigma associated with disability can cause exclusion, harassment, and bullying. Disability stigma and discrimination is compounded when an astronomer identifies with multiply marginalized identities; ending the erasure of disabled people of color needs to a central focus in accessibility and inclusion work[37].

Historically, the academic field of Disability Studies has focused on physical and sensory disabilities leading to workplace and institutional policies designed to provide accommodations in these areas. This focus on the physical and sensory leaves out an important area: the mental and intellectual. The current academic culture is at odds with acknowledging and supporting mental disability and neurodiversity. Often, those with mental disabilities or who are neurodiverse leave the field or are forced out along the way, but even those who stay can feel isolated and unwelcome[14]. Additionally, intersectional approaches are often not considered: microaggressions[38,39,40], aggression, overt racism, and disparagement of gender and sexual minority identities can negatively impact the mental health of astronomers of color and LGBTQIA+ members of our communities[41,42].

Due to societal stigma, many people do not feel comfortable disclosing their mental disability or neurodiversity and thus do not seek accommodations. When accommodations are available they often require the individual to publicly out themselves with a formal diagnosis. As formal diagnoses are expensive and not always based on inclusive criteria[43,44,45], individuals with multiply marginalized identities may not be able to access needed accommodations. Colleges and universities often focus mental health outreach efforts on students, not faculty or staff. This lack of advertising combined with the cultural stigma against mental disabilities, often leaves faculty and staff feeling unsupported[15]. In general, the current academic culture of astronomy causes and contributes to degradation of mental health, by prizing success at all costs and often celebrating 80-100 hour work weeks, loss of sleep, lack of eating, or poor self-care[16].

While we have listed above some specific barriers, recommending a checklist of solutions unique to those barriers is not general or flexible enough to anticipate the needs of astronomers before they have had to ask for support or accommodations.

# IV. Moving forward toward a truly accessible astronomy

The barriers to access we have outlined above all serve to exclude the participation of astronomers and astronomy students with disabilities, especially with congenital and early onset disabilities. We envision a future when the work done toward inclusivity brings accessibility and disability into the mainstream. By "mainstream," we mean when accessibility is a 0th order consideration in the development of any new facility, database, or software product; when accessibility is highest priority in planning meetings, or assessing publications and related tools. Being mainstream would include giving work promoting inclusivity respect and recognition deserved for the critical role it plays in creating a culture that welcomes and supports the people science needs; inclusion work should be prestigious, and not an afterthought or another box to check off. In an inclusive field, astronomers with disabilities would be able to trust that an unsuccessful job or funding application was based upon a fair evaluation and not because we may be considered a burden to the institution or funding agency. When accessibility is mainstream, universal design is the default, and astronomers with disabilities could focus on their science rather than worry about whether they will be able to access information and resources. Below, we outline specific ways that accessibility can be brought to the forefront: specific ways to better quantify the losses to the field due to inaccessibility, ideas for developing institutional supports, and incentives to accelerate the process.

## IV-A. Astronomers and professional organizations

Professional organizations for astronomers provide myriad valuable services to their membership, including organizing meetings, owning and managing publications, providing avenues for communicating current issues to the membership, and even issuing statements and calling for broader societal change that benefits science and scientists. We look to the AAS primarily for leadership and help in accessibility and inclusion and to set examples and precedent for all professional organizations representing astronomers. Below, we summarize examples of best practices for astronomical organizations to ensure the services they provide are fully accessible and inclusive. Additionally, we encourage flexibility, receptiveness to feedback, and a dynamic, proactive approach to inclusion that continues to rise to meet and extend goals even as progress is made.

Recommendations for promoting inclusivity in professional astronomical societies:
- Provide mechanisms for protection and, if needed, recourse, for junior scientists who are discriminated against, harassed, bullied, or experience violations of academic integrity.
    - Work with inclusion groups (in the AAS, these are the Committee on the Status of Women in Astronomy[46], the Committee on the Status of Minorities in Astronomy[47], the Committee for Sexual-orientation and Gender Minorities in Astronomy[48], and WGAD) to ensure compliance with Titles VI and IX.
- Adopt editorial policies that ensure research publications and data sets are highly accessible. Establish accessibility guidelines for authors. Seek feedback from readers with disabilities.
- Peer review of papers should be based on a metric to prevent harassment, insults, and bullying. The peer reviewers should not be anonymous.
- Require all databases to be ADA-compliant and accessible.
- Recognize inclusion work as prestigious and important. Establish a board to identify individuals and set up a high profile award system to motivate the work for inclusion and recognize those doing dignifying work.
- Encourage funding agencies to pursue full-field surveys and be transparent with all demographic data related to funded proposals' participants.

Recommendations for ensuring that meetings are accessible and inclusive:
- Designate a meeting accessibility lead who is an employee of the professional society. The accessibility lead should work proactively with presenters, attendees and organizers to provide the most accessible experience possible.
- Endorse and adopt best practice recommendations on meetings accessibility (within the AAS, the WGAD is preparing a document for this purpose; Monkiewicz et al. *in prep*) and ensure any division or topical meeting sponsored by the organization follows the meeting accessibility guidelines.
- Require that funds be set aside in advance for real-time captioning and/or American Sign Language (ASL) interpretation, rather than attempting to be made available only if on demand.
- After the conclusion of the meeting solicit anonymous feedback from attendees on what was done well and what can be improved.

## IV-B. Institutions and academic departments

Accessibility barriers at astronomy departments and institutions come in forms such as building infrastructure, educational practices, and culture[49]. Many astronomical facilities that pre-date the ADA contain wheelchair inaccessible areas and other structural barriers. Departments should strive for 100% ADA compliance, while recognizing that ADA guidelines are minimum standards. The following list includes several ways that institutions can become more accessible.
- Proactively improve accessibility for people with disabilities.
  - Perform accessibility audits and/or host AAS Climate Site Visits[50].
  - Use the results to develop and enact an accessibility roadmap.
  - Establish design standards and specifications for new construction and renovations (beyond the ADA) that take into account local topography, climate, and architecture.
- Create accessible learning environments.
  - Practice universal design when developing course materials, electronic media, equipment, and research products[39].
  - Provide multiple modes of access, communication, and evaluation.
  - Include accessibility information on webpages and course syllabi.
  - Provide accurate closed captions on videos and alt-text for graphics.
- Create accessible working environments.
  - Provide office spaces which are friendly to those with sensory processing issues, regardless of formal diagnosis or public acknowledgement. Examples include minimizing the use of open plan office spaces, providing dedicated rooms for telecons and meetings, and allowing for "non-social" shared offices.
  - Include accommodation request information on job listings. Assure applicants that any requests will not be known to the search committee or held against the applicant.
  - Prioritize accessibility when planning and hosting events.
- Develop a culture of access[51].
  - Respect disability accommodation requests.
  - Engage with institutional accessibility services to be prepared in advance rather than responding to accommodation requests on the fly.
  - Recognize the importance of effective mentoring in student success, and connect students to virtual mentoring services[52] or e-mentoring[53] if mentoring by a more senior disabled astronomer is unavailable.

## IV-C. From the top: funding agencies

Beyond professional societies and individual institutions, funding agencies can play an important role in incentivizing, rewarding, and encouraging progress beyond bare-bones compliance with the ADA. Where societies and institutions are unable to provide support, funding agencies can step in and provide resources as well as motivation for change. Below, we outline a few high-level recommendations of leadership that can create substantial positive change to quantify the losses to science we are currently experiencing due to inaccessibility, to understand the career trajectories and experiences of scientists with disabilities, and to increase awareness and will to progress toward an inclusive academia.

Demographic analysis recommendations:
- Surveying and collation of data to track presence and progress of scientists and scientists-in-training across their academic trajectory is needed to quantify representation at all career stages and identify major loss points.
- Surveying needs to be done by an independent party with input and oversight by professional astronomers, and needs to go beyond grant P.I. demographics or professional society memberships, as these two groups are not fully inclusive of all astronomers in the US.
- Data aggregation and analysis must go beyond a simple tally to capture the quality of experiences of people with disabilities in the field; there is a distinct gap between institutional reports of achievement and the reality of the scientists' experiences. This work can emulate the UN Commission on the Rights of People with Disabilities [54].

Funding proposal recommendations:
- Include accessibility as an important feature when evaluating proposals.
  - Require proposing institutions to provide accessibility statements that identity the current accessibility status of their facilities and future plans for increasing access. Hold institutions that have many unaddressed accessibility barriers accountable.
  - Ensure astronomers with disabilities are awarded grants with the same probability of success as able, neurotypical astronomers.
  - Transparency is needed in rates of funding for projects relative to aspects of identity, including disability, and these data should include all people providing effort on the grant, not just the P.I.'s.

Broader, general recommendations:
- Provide small grant opportunities to fund accessibility at meetings run by organizations with limited resources. Big organizations that receive funding for meetings should be held accountable for making events accessible.
- Incentivize non-AAS publishers to play a role in publication accessibility. Involve astronomers with disabilities in the process.
- Provide support for US-based groups working for accessibility and inclusion to coordinate with international groups working toward the same goals.

https://www.insidehighered.com/blogs/gradhacker/disabled-grad-school-i-too-dread-accom

modations-talk. (Accessed: 10th July 2019)

21. Disabled in Grad School: Informal Accommodations | Inside Higher Ed. Available at:

    https://www.insidehighered.com/blogs/gradhacker/disabled-grad-school-informal-accommod

    ations. (Accessed: 10th July 2019)

22. Winning: A College Guide for Students with Disabilities. *EDsmart* (2016).

23. Barazandeh, G. & Sereseres, C. D. Attitudes Toward Disabilities and Reasonable

    Accommodations at the University. 12

24. Listening to the patterns of the universe | EarthSky.org. Available at:

    https://earthsky.org/space/space-data-into-sound-patterns-wanda-diaz-merced. (Accessed:

    9th July 2019)

25. Diaz-Merced, W. L. *et al.* Sonification of Astronomical Data. in **285**, 133–136 (2012).

26. Hahn, H. Toward a politics of disability: Definitions, disciplines, and policies. *Soc. Sci. J.* **22**,

    87–105 (1985).

27. Roweton, W. E. Linton, S. (1998). Claiming disability: Knowledge and identity. New York:

    New York University Press. 203 pp., $16.95. *Psychol. Sch.* **37**, 557–558 (2000).

28. Owens, J. Exploring the critiques of the social model of disability: the transformative

    possibility of Arendt's notion of power. *Sociol. Health Illn.* **37**, 385–403 (2015).

29. Gustavsson, A. The role of theory in disability research –springboard or strait–jacket?

    *Scand. J. Disabil. Res.* **6**, 55–70 (2004).

30. Disability Studies A Nordic Perspective. *studylib.net* Available at:

    https://studylib.net/doc/9596800/disability-studies-a-nordic-perspective. (Accessed: 10th

    July 2019)

31. Tøssebro, J. Introduction to the special issue: Understanding disability. *Scand. J. Disabil.*
13